# Global Representation of the Fine Structure Constant and its Variation

Michael E. Tobar

Frequency Standards and Metrology Research Group, School of Physics, M013, University of Western Australia Crawley, 6009, WA, Australia

**Abstract**

The fine structure constant, $\alpha$, is shown to be proportional to the ratio of the quanta of electric and magnetic flux of force of the electron, and provides a new representation, which is global across all unit systems. Consequently, a variation in $\alpha$ was shown to manifest due to a differential change in the fraction of the quanta of electric and magnetic flux of force, while a variation in $\hbar c$ was shown to manifest due to the common mode change. The representation is discussed with respect to the running of the fine structure constant at high energies (small distances), and a putative temporal drift. It is shown that the running of the fine structure constant is due to equal components of electric screening (polarization of vacuum) and magnetic anti-screening (magnetization of vacuum), which cause the perceived quanta of electric charge to increase at small distances, while the magnetic flux quanta decreases. This introduces the concept of the 'bare magnetic flux quanta' as well as the 'bare electric charge'. With regards to temporal drift, it is confirmed that it is impossible to determine which fundamental constant is varying if $\alpha$ varies.

PACS: 04.80.Cc, 06.20.Jr

Keywords: Fundamental Constants, Fine Structure Constant, Magnetic Flux Quanta

**1. Introduction**

The fine structure constant is very important in QED [1], It was first introduced by Sommerfeld [2] to explain the fine structure in hydrogen atoms, as the ratio of the electron velocity to the speed of light; the meaning of the constant has now evolved to be a measure of the strength of the electromagnetic field. Recent experimental evidence that the fine structure constant may be drifting [3,4] and has triggered much interest in theories that account for the drift in fundamental constants [5-10]. Also, it has provided stimulus to laboratory tests, which aim to improve the precision of measurements of the constancy of the fine structure constant [11-13]. Furthermore, it is well established that the fine structure constant varies with energy (or distance) as one probes close to the electron [14]. In this paper the fine structure constant is represented in terms of only electromagnetic quantities[15,16]. For the SI unit representation, this includes the quantum of electric charge $e$, the quantum of magnetic flux $\phi_0$, and the permittivity and permeability of free space, $\varepsilon_0$ and $\mu_0$.

To solve systems in Classical Electrodynamics it is common to represent the problem in terms of the co-ordinates of charge or magnetic flux. Both are conserved quantities and are considered dual variables[17], and one may formulate a problem in terms of either of these quantities and get the same solution. This fact may lead one to consider solving quantum systems in terms of magnetic flux, which has been attempted in the past. Jehle spent a large part of his life developing a theory of the electron and elementary particles based on quantized magnetic flux loops[18]. Also, Dirac suggested the existence of the magnetic monopole to describe the charge of the electron[19]. Both these theories rely on a physical relationship between flux and charge. In the case of the Jehle model, he suggested that the electron was made of quantized flux loops, spinning at the Zitterbewegung frequency. Since it is well known the magnetic flux is quantized, it seems plausible that the quantized value of flux should play a role in the description of particle physics. Jehle developed a series of papers, which attempted to do this based on the quantized flux loop[18,20-22]. The relation between the electric and magnetic properties is fundamental to electrodynamics as charge in motion produces magnetic flux.

**2. Alternative representation of the fine structure constant**

Usually the fine structure constant (or coupling constant) is defined from the static Coulomb force between two charges. The force, $F_e$, in SI units is given by;

$$F_e = \frac{e^2}{4\pi\varepsilon_0 r^2} \tag{1}$$

Here $e$ is the electric charge in Coulombs, $r$ is the separation between the charges in meters, and $\varepsilon_0$ is the permittivity of free space in Farads per meter. To consider the strength between two static charges in terms of fundamental constants, one must ignore the inverse square nature of the force and just consider the constant of proportionality $e^2/4\pi\varepsilon_0$. Because this proportionality constant has the dimensions of energy × distance, the dimensionless constant can be constructed by dividing by $\hbar c$, which also has the same units. Thus, it is usual to write the fine structure constant in SI units as the following;

$$\alpha = \frac{e^2}{4\pi\varepsilon_0 \hbar c} \tag{2}$$

Similar coupling constants are also written for the strong, weak and gravitational forces. If one analyses these coupling constants it may be seen that they too are represented as a dimensionless constant by comparing the constants of proportionality of the force with $\hbar c$ [1]. Thus, when we discuss these dimensionless constants relative to one another they represent comparative strengths of the different forces. For example, the strong force coupling constant, is approximately one, the electromagnetic is 1/137 the weak is $10^{-6}$, and gravity $10^{-39}$.

The logic of defining (2) above, leads many to state that $\alpha$ must be proportional to the strength of the Coulomb (electric) force, as it is proportional to $e^2$. However, atomic systems are not just comprised of static charge, as they also exhibit spin and magnetic moments. Thus, one may expect the magnetic nature to be present in the definition as well as the electric. The magnetic nature is actually hidden in the $\hbar c$ term that we divided by. For example, all transitions between electron orbit and spin states, when they interact with electromagnetic radiation, are governed by the following equation

$$E_{ph} = \frac{\hbar c}{\lambda_{ph}} \qquad (3)$$

where $E_{ph}$ is the energy of the absorbed or emitted photon, $\lambda_{ph}$ is the wavelength and $\hbar c$ is the constant of proportionality. Also, $\hbar c$ is the proportionality constant that relates the Casimir force to the dimensions of two electromagnetic neutral electrodes, and thus incorporates both the electric and magnetic zero point properties of the vacuum.

The key relations between the fundamental constants for classical and quantum electromagnetism in SI units are:

$$c = \frac{1}{\sqrt{\varepsilon_o \mu_o}} \qquad (4)$$

$$\hbar = \frac{e\phi_0}{\pi} \qquad (5)$$

Here $\mu_0$ is the permeability of free space and $\phi_0$ is the quantum of magnetic flux due to the spin of the electron. The flux is equal to the minimum quanta of spin angular momentum, divided by the quanta of charge, and is the same as the flux produced by a Cooper pair[23-25] (This was recently derived in[26] on the basis of the magnetic top model). Equation (4) and (5) basically give a simple description of the relation between electric and magnetic quantities. Now given that the fine structure constant is an electromagnetic constant it would be instructive to substitute (4) and (5) into (2) to actually express it in terms of the electric and magnetic constants, $\phi_0$, $e$, $\varepsilon_0$ and $\mu_0$. If we do this (2) becomes:

$$\alpha = \frac{1}{4}\frac{e}{\phi_o}\sqrt{\frac{\mu_o}{\varepsilon_o}} \qquad (6)$$

This representation of the fine structure constant is now expressed as ratios of the classical and quantum electromagnetic constants. In actual fact $\sqrt{\mu_o/\varepsilon_o}$ is the impedance of free space and $e/2\phi_o$ is the Quantum Hall conductance.

**2.1 Representation in terms of static magnetic and electric flux of force**

In this section a representation of the fine structure constant is formulated in terms of the quanta of static magnetic and electric flux of force, which turns out to be global for all unit systems (see appendix A). First, a simple classical static model of a magnetic flux is introduced, which is similar to a magnetic circuit. Since magnetic fields are solenoidal, magnetic flux may be modeled as a loop (i.e. no matter on how complicated the path it will eventually come back to where it started). A simple circuit model of this type is shown in figure 1. One may consider the flux loop as a circular magnet with a north and a south pole held together by an attractive magnetic force. In SI units the force between the north and south pole is given by;

$$F_\phi = \frac{\phi_o^2}{2\mu_o A} \tag{7}$$

where $A$ is the effective cross section area of the magnetic loop, as shown in figure 1.

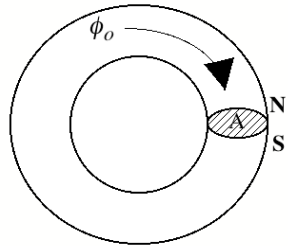

Figure 1. Schematic of a magnetic flux loop with effective cross section area A.

The magnetic and Coulomb force given by equations (1) and (7) are not constants. For example, (1) follows the inverse square law and (7) depends on the cross section area over the path of the magnetic flux loop. However, the flux of electric and magnetic force, defined as the force field multiplied by the cross sectional area perpendicular to the field lines, are constants and in SI units are given by

$$\Phi_e = \frac{e^2}{\varepsilon_o} \quad \text{and} \quad \Phi_\phi = \frac{\phi_o^2}{2\mu_o} \tag{8}$$

Here, $\Phi_e$ and $\Phi_\phi$ may be considered as the flux of electric and magnetic force generated by a static quantized charge, $e$, and a static quantized magnetic flux loop, $\phi_0$, respectively.

Considering equation (4), (5), (6) and (8) one can then show the following relations;

$$\alpha = \frac{1}{4\sqrt{2}} \sqrt{\frac{\Phi_e}{\Phi_\phi}} \qquad (9)$$

$$\hbar c = \frac{\sqrt{2}}{\pi} \sqrt{\Phi_e \Phi_\phi} \qquad (10)$$

Equations (9) and (10) together now portray a representation of $\alpha$ and $\hbar c$, which are symmetric in its electric and magnetic parts.

In appendix A it is shown that the equations (9) and (10) are global representations for all unit systems, which is important if one wants to analyze the variation of $\alpha$ in terms of dimensioned constants. In the following section, this permits an analysis of the running of the fine structure constant, as well as putative temporal variations, which would not be otherwise possible.

## 3. Variation of the fine structure constant in terms of electric and magnetic quantities

It is interesting to consider the meaning of (9) and (10) if the Fine Structure Constant varies. Because they are independent of unit representation they may be implicitly differentiated to obtain;

$$\frac{\Delta \alpha}{\alpha} = \frac{1}{2}\left(\frac{\Delta \Phi_e}{\Phi_e} - \frac{\Delta \Phi_\phi}{\Phi_\phi}\right) \qquad \frac{\Delta(\hbar c)}{\hbar c} = \frac{1}{2}\left(\frac{\Delta \Phi_e}{\Phi_e} + \frac{\Delta \Phi_\phi}{\Phi_\phi}\right) \qquad (11)$$

Thus, we have succeeded in representing a change in $\alpha$ and $\hbar c$ in terms of the change of the electric and magnetic quantities defined by (8). Here a change in $\alpha$ is due to a differential variation in the strength of the electric and magnetic force, while a change in $\hbar c$ is due to a common mode variation.

### 3.1 The running electromagnetic coupling constant

It is well established that at high energies as one probes closer to the sub-structure of an electron, the fine structure constant increases[14]. This is due to the perceived screening of the bare electron charge by polarized virtual electrons at low energies, E (or large distance). If (11) is applied to this situation, we can assume $\hbar c$ does not vary and the following must be true,

$$\frac{1}{\hbar c}\frac{\partial(\hbar c)}{\partial E} = 0 \qquad \frac{1}{\Phi_e}\frac{\partial \Phi_e}{\partial E} = -\frac{1}{\Phi_\phi}\frac{\partial \Phi_\phi}{\partial E} \qquad (12)$$

and thus,

$$\frac{1}{\alpha}\frac{\partial \alpha}{\partial E} = \frac{1}{\Phi_e}\frac{\partial \Phi_e}{\partial E} = -\frac{1}{\Phi_\phi}\frac{\partial \Phi_\phi}{\partial E} \qquad (13)$$

Thus at low energies, the density of the curl-free electric lines of force is reduced, due to the vacuum field opposing the field of the bare electron. In actual fact the virtual electrons are also in motion and must posses magnetic properties as well[27]. In contrast at low energies, the density of the divergence-free magnetic lines of force of the bare electron is enhanced. This means the magnetic moments of the virtual electrons align with the magnetic moment of the bare electron, and one can say the vacuum is also magnetized like a paramagnet. This has the effect of the magnetic flux quanta reducing to a lower value at high energies, in the opposite way to the charge, and the magnetic flux quanta is thus anti-screened. This leads to the concept of the "bare magnetic flux-quanta" being less than the magnetic flux quanta at low energies. Previously[28], this concept has also been explained by a 'bare magnetic monopole charge'. However, we point out here that it is not necessary to introduce the concept of the magnetic monopole to describe this phenomenon, when one may define it in terms of closed loops of flux. Also, because of equation (11) it is evident that the running of the fine structure constant is due to equal components of electric screening and magnetic anti-screening.

**3.2 Putative drift of the fine structure constant**

It is widely accepted that when considering problems that deal with time variations of fundamental constants, that it only makes sense to consider dimensionless constants[7,10,29-33]. However, because (9) and (11) are global across all unit systems, a putative drift in the fine structure constant may be interpreted as a differential drift between the strength of the electric and magnetic flux of force (which of course have dimension). In this section we show that this finding does not contradict the accepted belief, that one cannot determine whether or not $e$ or $c$ drifts if there is a putative drift in $\alpha$[7,10,29-33]. In general one could also consider $\hbar$ as well as $e$ or $c$, but mostly it has been assumed to remain constant. In this work we consider more generally the product $\hbar c$, as it naturally fits with the magnetic and electric flux of force representation.

Case I: Firstly, it is assumed that the total electromagnetic energy of an electron remains constant. In this case, $\Delta\Phi_e/\Phi_e = -\Delta\Phi_\phi/\Phi_\phi$, and from (11) the putative drift in $\alpha$ may be written as;

$$\frac{\Delta\alpha}{\alpha} = \frac{\Delta\Phi_e}{\Phi_e} = -\frac{\Delta\Phi_\phi}{\Phi_\phi} = 2\frac{\Delta e}{e} \quad \text{and} \quad \frac{\Delta(\hbar c)}{\hbar c} = 0 \quad (14)$$

Physically this means that the electric energy would be converted to magnetic energy or vice versa. If this occurred one would then get a drift in $\alpha$ independent of $\hbar c$ due to only a differential change in the electric and magnetic flux of force.

Case II: Secondly, the magnetic and electric parts are assumed to vary at the same rate (common mode drift). In this case $\Delta\Phi_e/\Phi_e = \Delta\Phi_\phi/\Phi_\phi$, and from (11) $\alpha$ will not drift but $\hbar c$ will. In this case another energy process would need to be involved.

Case III: Finally, the case is considered when putative $\alpha$ and $\hbar c$ drift are related. In this case, the differential and common mode components of (11) must be correlated. This would occur if, for example, another form of energy was converted to only magnetic or electric energy but not both, such that either $\Delta\Phi_e/\Phi_e = 0$ or $\Delta\Phi_\phi/\Phi_\phi = 0$. In this case;

$$\frac{\Delta\alpha}{\alpha} = \frac{\Delta(\hbar c)}{\hbar c} \text{ if } \frac{\Delta\Phi_\phi}{\Phi_\phi} = 0, \text{ and } \frac{\Delta\alpha}{\alpha} = -\frac{\Delta(\hbar c)}{\hbar c} \text{ if } \frac{\Delta\Phi_e}{\Phi_e} = 0, \text{ with } \frac{\Delta e}{e} = 0 \qquad (15)$$

Thus, to unequivocally interpret which fundamental constant drifts if $\alpha$ drifts, both the common mode component and differential component of (11) must be known. However since $\hbar c$ (common mode component) has dimension it makes no sense to try and measure its drift. Actually, the measurement of drift in $\hbar c$ has been attempted previously, for reviews on these measurements see [1]. However, Bekenstein[34] showed these attempts generated null results as the constancy was actually implied in the analysis. Therefore, we still may conclude that the absolute variation of any of the dimensioned fundamental constant, whether it be $e$, $\hbar c$, $\Phi_e$ or $\Phi_\phi$, cannot be measured. However, despite this, the time variation of $\alpha$ may be interpreted as the differential drift of the strength of the Coulomb force with respect to the magnetic force, even though they have dimension.

4. Discussion

The fine structure constant, $\alpha$, may be represented in many ways depending on the unit system and the selection of fundamental constants. In this paper, a representation, which is global across all unit systems, has been successfully obtained in terms of the quanta of electric and magnetic flux of force of the electron. It was shown that $\alpha$ is proportional to the ratio of the square root of quantized electric and magnetic flux of force, while $\hbar c$ was shown to be proportional to the product. Because the representation is global across all unit systems an analysis of the variation of the fine structure constant was successfully made. The variation in $\alpha$ was described as a differential change in the flux of force associated with the quanta of electric charge and magnetic flux of the electron. In contrast, a change in $\hbar c$ (and hence Casimir Force) can be described as a common mode change in the same variables. With regards to the running of $\alpha$ at high energies, it was shown that it may be described by both vacuum polarization and magnetization effects, and introduces the new concept of the 'bare magnetic flux quanta'. It was also shown that it is not possible to determine the physical process behind any putative $\alpha$ drift, as the measurement process does not allow the determination of the common mode drift of the quanta of electric and magnetic flux of force, even if it occurs.

**Appendix. Global unit system**

In this appendix equations (9) and (10) are shown to be global across all unit systems by converting to the generalized unit system of Jackson[35]. For this unit system Maxwell's equations in vacuum are given by[35]

$$\nabla \cdot \boldsymbol{E}_g = 4\pi k_1 \rho_g \; ; \; \nabla \cdot \boldsymbol{B}_g = 0 \; ; \; \nabla \times \boldsymbol{B}_g = 4\pi \frac{k_2}{k_3} \boldsymbol{J}_g + \frac{k_2}{k_1 k_3} \frac{\partial \boldsymbol{E}_g}{\partial t} \; ; \; \nabla \times \boldsymbol{E}_g = -k_3 \frac{\partial \boldsymbol{B}_g}{\partial t} \quad (A1)$$

Here the subscript $g$ refers to the general units of Jackson (which encompasses all electromagnetic unit systems depending on the values of $k_1$, $k_2$ and $k_3$). The conversion between SI and Jackson's units can be simply made with the following substitutions[35]:

$$\frac{k_2}{k_3} - > \frac{\mu_0}{4\pi} \; , \; k_1 - > \frac{1}{4\pi\varepsilon_0}, \; c_g^2 k_3 - > \frac{1}{\mu_0 \varepsilon_0} \quad (A2)$$

Thus in the general unit system (6) becomes,

$$\alpha = \pi \frac{e_g}{\phi_{0,g}} \sqrt{\frac{k_2 k_1}{k_3}} \quad (A3)$$

where the vacuum impedance is given by $Z_{0,g} = 4\pi\sqrt{k_2 k_1 / k_3}$ and the quantum Hall resistance by $R_{h,g} = 2\phi_{0,g} / e_g$. In the same way, the electric, $\Phi_{e,g}$, and magnetic, $\Phi_{\phi,g}$, flux of force of the electron given by (8) can be shown to be

$$\Phi_{e,g} = 4\pi k_1 e_g^2 \; \text{and} \; \Phi_{\phi,g} = \frac{1}{8\pi} \frac{k_3}{k_2} \phi_{0,g}^2 \quad (A4)$$

Combining (A4) with (A3) to eliminate $k_1$, $k_2$ and $k_3$ gives;

$$\alpha = \frac{1}{4\sqrt{2}} \sqrt{\frac{\Phi_{e,g}}{\Phi_{\phi,g}}} \quad (A5)$$

which is equivalent to (9) as expected.

To proceed and show that (10) is also global across all unit systems, the fine structure constant given in (2) can be represented in the global unit system as:

$$\alpha = k_1 \frac{e_g^2}{\hbar_g c_g} \quad (A6)$$

Combining (A4), (A5) and (A6) to eliminate $k_1$ and $\alpha$, the following is obtained

$$\hbar_g c_g = \frac{\sqrt{2}}{\pi}\sqrt{\Phi_{e,g}\Phi_{\phi,g}} \tag{A7}$$

as expected.

From the above analysis, it is also possible to represent the constants $k_1$, $k_2$ and $k_3$ in terms of only fundamental constants

$$k_1 = \frac{\alpha\,\hbar_g c_g}{e_g^2},\ k_2 = \frac{\alpha\,\hbar_g}{e_g^2 c_g},\ k_3 = \frac{\pi^2\,\hbar_g^2}{\phi_{0,g}^2 e_g^2} \tag{A8}$$

and is valid for all unit systems. Examples for four commonly chosen unit systems are shown below in table I.

To uniquely determine an arbitrary unit system in classical electrodynamics and in vacuum, one needs to specify the three constants $k_1$, $k_2$ and $k_3$.[†] However, in quantum electrodynamics one other constant that specifies the quantum nature must also be specified, along with $k_1$, $k_2$ and $k_3$. For example, one could choose one of $\hbar_g$, $e_g$ or $\phi_{0,g}$ and all others would follow by applying equations (A1) to (A8). Another approach would be simply to choose the values of the four fundamental constants $c_g$, $\hbar_g$, $e_g$ and $\phi_{0,g}$, then all other parameters in the unit system including $k_1$, $k_2$ and $k_3$, could be determined.

Table 1. Values of some constants of quantum electrodynamics for some selected unit systems.

| Global Parameter | SI Units | Natural Units | Rydberg Units | CGS Units |
|---|---|---|---|---|
| $k_1$ | $10^{-7} c_{SI}^2$ | 1 | 1 | 1 |
| $k_2$ | $10^{-7}$ | 1 | $\alpha^2/4$ | $1/c_{CGS}^2$ |
| $k_3$ | 1 | 1 | $\alpha/2$ | $1/c_{CGS}$ |
| $\hbar_g$ | $10^{-7} e_{SI}^2 c_{SI}/\alpha$ | 1 | 1 | $e_{CGS}^2/c_{CGS}\alpha$ |
| $e_g$ | $e_{SI}$ | $\sqrt{\alpha}$ | $\sqrt{2}$ | $e_{CGS}$ |
| $\phi_{0,g}$ | $10^{-7}\pi e_{SI} c_{SI}/\alpha$ | $\pi/\sqrt{\alpha}$ | $\pi/\sqrt{\alpha}$ | $\pi e_{CGS}/\alpha\sqrt{c_{CGS}}$ |
| $c_g$ | $c_{SI}$ | 1 | $2/\alpha$ | $c_{CGS}$ |
| $Z_{0,g}$ | $10^{-7} 4\pi c_{SI}$ | $4\pi$ | $2\pi\alpha$ | $4\pi/\sqrt{c_{CGS}}$ |
| $R_{h,g}$ | $10^{-7} 2\pi c_{SI}/\alpha$ | $2\pi/\alpha$ | $\pi$ | $2\pi/\alpha\sqrt{c_{CGS}}$ |
| $\Phi_{e,g}$ | $10^{-7} 4\pi e_{SI}^2 c_{SI}^2$ | $4\pi\alpha$ | $8\pi$ | $4\pi e_{CGS}^2$ |
| $\Phi_{\phi,g}$ | $10^{-7} \pi e_{SI}^2 c_{SI}^2/8\alpha^2$ | $\pi/8\alpha$ | $\pi/4\alpha^2$ | $\pi e_{CGS}^2/8\alpha^2$ |

---

[†] The inclusion of media is left out as it adds the extra complication of including polarization and magnetization of magnetic and dielectric media without adding to the discussion. However, if one wants to define the permittivity and permeability within the unit system, it is necessary to consider these details (see Jackson for further details[35]).

Of the unit systems presented in Table I, Natural units are perhaps the simplest as $k_2$, $k_3$, $c_g$ and $\hbar_g$ are selected to be unity. In contrast, SI and CGS units were developed within a framework that would facilitate relating the standard units of mechanics to electromagnetism. In the SI system, the definition of the absolute ampere and the speed of light determine the parameters $c_{SI}$ = 299792458 m/s, $e_{SI}$ = 1.60217733×10$^{-19}$ C, $k_2$ = 10$^{-7}$ and $k_3$ = 1. The rest of the parameters listed in Table 1 can then be determined from these four values. In CGS units a similar approach could be made, with $c_{CGS}$ = 29979245800 cm/s, $e_{CGS}$ = 4.80320680×10$^{-10}$ statcoulombs, $k_2$ = $1/c_{CGS}^2$ and $k_3$ = $1/c_{CGS}$. Rydberg units are similar to CGS, with the speed of light and charge of the electron specified differently, $c_{CGS} = 2/\alpha$ and $e_{CGS} = \sqrt{2}$.

**Acknowledgements**

The author has had many interesting discussions with Dr. Ian Mc Arthur and Dr. Paul Abbott regarding this work. Also the author would like to thank Prof. Friedrich Hehl for critical assessment of the manuscript. This work was funded by the Australian Research Council.

**References**

1. J.-P. Uzan, Rev. of Mod. Phys. (2003).
2. A. Sommerfeld, Ann. Phys., Lpz., **41**, 1 (1916).
3. J. K. Webb, *et al.*, Phys. Rev. Lett. **82**, 884-887 (1999).
4. J. K. Webb, *et al.*, Phys. Rev. Lett. **87**, 091301 (2001).
5. J. D. Bekenstein, Phys. Rev. D. **66** (2002).
6. T. Damour, F. Piazza, and G. Veneziano, Phys. Rev. Lett. **89**, 081601 (2002).
7. P. C. Davies, T. M. Davis, and C. H. Lineweaver, Nature **418**, 602 (2002).
8. P. Forgacs, Z. Horvath, Gen.Rel.Grav.10:931-940, (1979).
9. P. Forgacs, Z. Horvath, Gen.Rel.Grav.11:205,1979.
10. J. Magueijo, Rep. Prog. Phys. 66 2025-2068, (2003).
11. J. D. Prestage, R. L. Tjoelker, and L. Maleki, Phys. Rev. Lett. **74**, 3511-3514 (1995).
12. T. Damour and F. Dyson, Nucl. Phys. B **480**, 37-54 (1996).
13. C. Salomon, N. Dimarcq, M. Abgrall, *et al.*, C. R. Acad. Sci. Paris **IV**, 1-17 (2001).
14. I. Levine et. al., Phys. Rev. Lett., **78** (3), pp. 424-427, (1997)
15. M.E Tobar, hep-ph/0306230
16. F.W. Hehl and Y.N. Obukhov, physics/0407022
17. F. W. Hehl and Y. N. Obukhov: Foundations of Classical Electrodynamics: Charge, Flux, and Metric. Birkhauser, Boston, MA, (2003)
18. H. Jehle, Phys. Rev. D. **3**, 306-345 (1971).
19. P. A. M. Dirac, Phys. Rev. **74**, 817-830 (1948).
20. H. Jehle, Phys. Rev. D. **6**, 441-457 (1972).
21. H. Jehle, Phys. Rev. D. **11**, 2147-2177 (1975).
22. H. Jehle, Phys. Rev. D. **15** (1977).
23. F. London, *Superfluids* (John Wiley, 1950).
24. L. Onsager, Proc. Int. Conf. Th. Phys., 935 (1953).
25. M Saglam, B Boyacioglu, Int. J. Mod. Phys. B, 16(4), 607-614 (2002).
26. B. S. Deaver and W. M. Fairbank, Phys. Rev. Lett. **7**, 43-46 (1961).
27. F. Wilczek, hep-th/9609099
28. A, F. Rañada and J. L. Trueba, Phys. Lett. B, 422 196-200 (1998).
29. M.J. Duff, hep-th/0208093.
30. G.F.R. Ellis, J-P. Uzan, gr-qc/0305099.
31. M.J. Duff, L.B. Okun, G. Veneziano, physics/0110060.
32. V.V. Flambaum, astro-ph/0208384.
33. H. B. Sandvik, J. D. Barrow, J. Magueijo, Phys. Rev. Let. 88, 031302 (2002).
34. J. D. Bekenstein, Comments Astrophys. **8**, 89 (1979).
35. J. D. Jackson, *Classical Electrodynamics*, Wiley, New York (1962).